\pdfoutput=1

\documentclass[11pt]{article}

\usepackage[preprint]{acl}

\usepackage{times}
\usepackage{latexsym}

\usepackage[T1]{fontenc}

\usepackage[utf8]{inputenc}

\usepackage{microtype}

\usepackage{inconsolata}

\usepackage{algorithm}
\usepackage{algorithmic}
\usepackage{multirow}
\usepackage{booktabs}
\usepackage{amsmath}
\usepackage{amssymb}
\usepackage{graphicx}
\usepackage{bbding}

\usepackage{newfloat}
\usepackage{listings}

\title{\textit{Synthetic Singers} \\
A Review of Deep-Learning-based Singing Voice Synthesis Approaches}

\author{
    Changhao Pan$^{1}$\footnotemark[1], 
    Dongyu Yao$^{1}$\thanks{$\,$ Equal Contribution.}, 
    Yu Zhang$^{1}$,
    \\
    \textbf{
    Wenxiang Guo$^{1}$, Jingyu Lu$^{1}$, Zhiyuan Zhu$^{1}$, Zhou Zhao$^{1}$\thanks{$\,$ Corresponding Author.}
    } \\
    $^{1}$Zhejiang University, Hangzhou, China \\
    \texttt{\{panch,dongyuyao,zhaozhou\}@zju.edu.cn}
}

\begin{document}
\maketitle
\begin{abstract}

Recent advances in singing voice synthesis (SVS) have attracted substantial attention from both academia and industry. With the advent of large language models and novel generative paradigms, producing controllable, high‑fidelity singing voices has become an attainable goal. Yet the field still lacks a comprehensive survey that systematically analyzes deep‑learning‑based singing voice synthesis systems and their enabling technologies.
To address the aforementioned issue, this survey first categorizes existing systems by task type and then organizes current architectures into two major paradigms: cascaded and end-to-end approaches. Moreover, we provide an in-depth analysis of core technologies, covering singing modeling and control techniques. Finally, we review relevant datasets, annotation tools, and evaluation benchmarks that support training and assessment. In appendix, we introduce training strategies and further discussion of SVS. This survey provides an up-to-date review of the literature on SVS models, which would be a useful reference for both researchers and engineers. Related materials are available at~\url{https://github.com/David-Pigeon/SyntheticSingers}.

\end{abstract}

\section{Introduction}

Singing voice synthesis (SVS) seeks to generate high‑fidelity singing from textual lyrics and symbolic musical scores, and has garnered sustained interest in both industry and academic communities. The field has evolved from early engines such as VOCALOID\footnote{\url{https://www.vocaloid.com/en/}} and virtual singers like Hatsune Miku\footnote{\url{https://vocaloid.fandom.com/wiki/Hatsune_Miku}} to contemporary song‑generation platforms, including Seed‑Music~\cite{bai2024seed} and Suno\footnote{\url{https://suno.com/home}}, which deliver immersive, realistic listening experiences. Unlike conventional text‑to‑speech, SVS accepts richer inputs and imposes stricter constraints: synthesized vocals must articulate lyrics clearly while strictly following the prescribed melody. Moreover, modern SVS systems are expected to preserve fidelity, provide controllable stylistic variation, and convey expressive emotion.

Driven by the development of deep learning~\cite{vaswani2017attention} and generative models~\cite{ho2020denoising}, recent years have witnessed remarkable advances in SVS. Relative to traditional methods such as waveform concatenation~\cite{kenmochi2007vocaloid} and statistical parametric synthesis~\cite{saino2006hmm}, deep‑learning approaches provide finer acoustic detail~\cite{chen2020hifisinger} and markedly enhance the clarity and naturalness of the generated vocals~\cite{zhang2022wesinger}. Benefiting from the strong extensibility of modern generative models, contemporary SVS systems further deliver superior controllability and generalization, enabling zero‑shot singing generation while consistently maintaining high audio quality~\cite{zhang2024tcsinger,byun2024midi}.

The evolution of singing‑voice synthesis begins with the canonical goal of producing high‑fidelity vocals. Early systems, largely inherited from text‑to‑speech frameworks~\cite{wang2017tacotron,ren2019fastspeech}, reach this objective by introducing music‑specific modules such as score encoders~\cite{lu2020xiaoicesing} and pitch predictors~\cite{he2023rmssinger}. As generative AI matures and vision‑domain applications proliferate~\cite{rombach2022high,peebles2023scalable}, users grow to expect equally personalized and controllable audio experiences. Researchers and engineers have therefore advanced multi‑speaker models capable of high‑quality timbre generation~\cite{cho2022mandarin,huang2021multi}, along with editing schemes that enable fine‑grained control over emotion and singing style~\cite{hong2023unisinger,dai2024expressivesinger}. The advent of multimodal large language models~\cite{achiam2023gpt,chu2024qwen2} has raised expectations for customized singing‑voice generation. Zero‑shot SVS systems endowed with style control and transfer~\cite{zhang2024tcsinger}, have therefore received considerable attention. Meanwhile, emerging tasks such as speech‑to‑singing conversion~\cite{li2023alignsts} and text‑to‑song generation~\cite{liu2025songgen,yuan2025yue} have gained momentum, markedly expanding the application landscape of SVS across the entertainment, education, and film.

On synthesis process, the continuity of audio signals and the inherently one‑to‑many mapping from score to waveform initially led researchers to adopt an \textit{“acoustic‑model + vocoder”} pipeline~\cite{lu2020xiaoicesing}. In this cascaded approach, an acoustic model predicts intermediate acoustic features from the score, after which a vocoder converts these features into the waveform. To reduce error accumulation and utilize the prior knowledge in LLMs, recent works move toward vocoder‑free frameworks that generate waveforms directly. We refer to these newer systems as end‑to‑end approaches. Building on prevailing SVS architectures, we identify three core technologies: singing‑voice modeling, control mechanisms, and training strategies.

Considering the aforementioned factors, the organization of this survey is shown below. Section~\ref{sec:task} introduces the mainstream tasks and scenarios of SVS. Section~\ref{sec:arch} examines the different architectures of SVS models. We discuss the method for modeling and control in Section~\ref{sec:modeling}. Finally, we present resources of SVS models in Section~\ref{sec:resources}.

\section{Tasks of SVS}\label{sec:task}

\begin{figure*}[ht]
\centering
\includegraphics[width=0.99\linewidth]{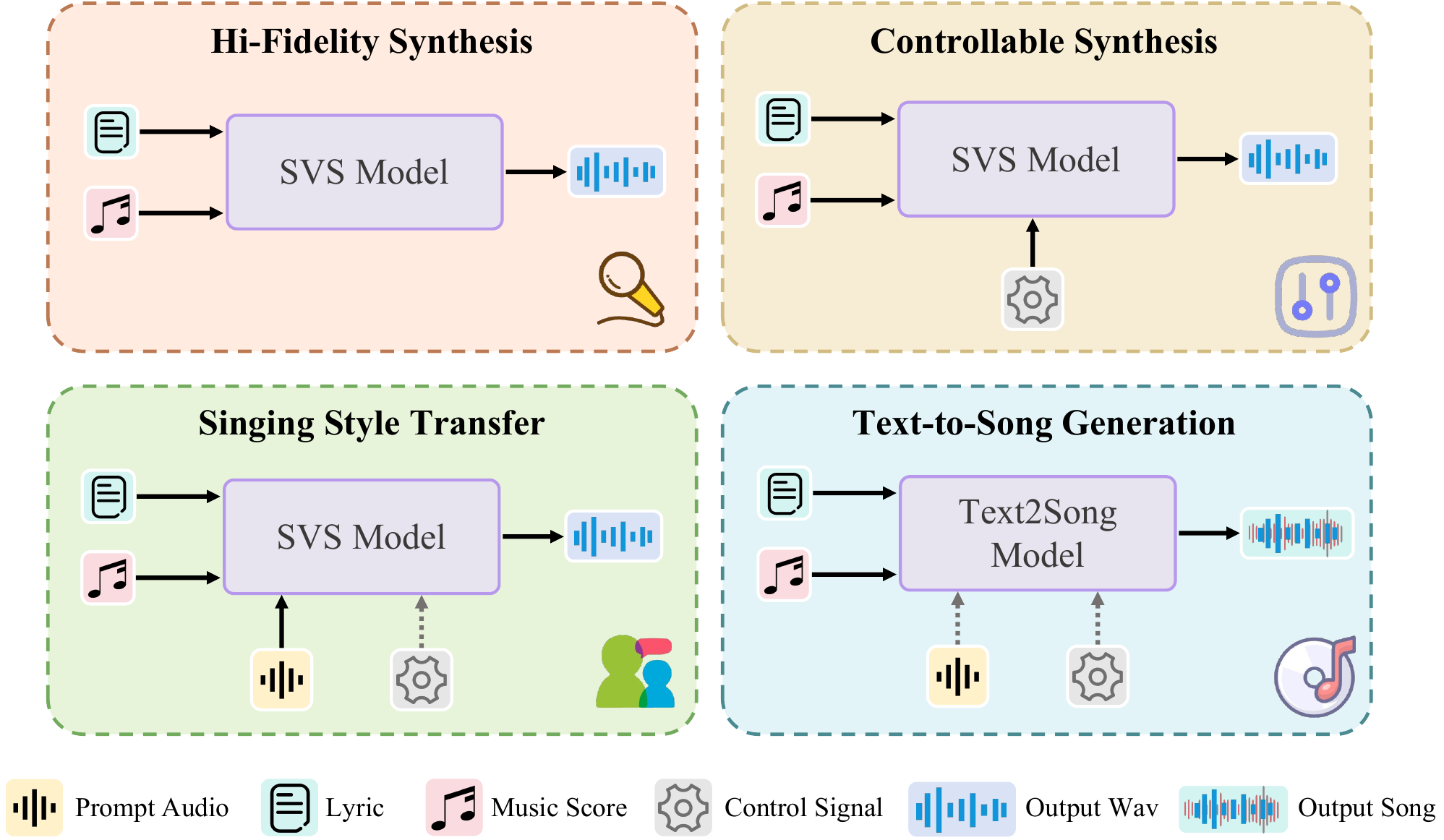}
\caption{
An overall demonstration of four prevailing tasks of SVS systems.
}
\label{fig:task}
\end{figure*}
\vspace{-5pt}

Drawing on demonstrations and user interfaces from SVS systems such as DiffSinger~\cite{liu2022diffsinger} and UTAU\footnote{\url{https://utau-synth.com}}, and referencing the historical progress of SVS, we divide the prevailing SVS tasks into four categories as shown in Figure~\ref{fig:task}.

\paragraph{Hi-Fidelity Synthesis}
High‑fidelity vocal reproduction is the foundational goal in SVS research~\cite{wagner2010experimental}. Early work pursued greater naturalness by embedding music‑specific inductive biases. XiaoiceSing~\cite{lu2020xiaoicesing} extends the FastSpeech architecture~\cite{ren2019fastspeech} with constraints on note duration and pitch, while HiFiSinger~\cite{chen2020hifisinger} replaces the conventional vocoder in XiaoiceSing with Parallel WaveGAN~\cite{yamamoto2020parallel}, achieving superior audio fidelity.
To counter the over‑smoothing typical of Transformer‑based generators, DiffSinger~\cite{liu2022diffsinger} introduces a shallow‑diffusion DDPM, markedly improving spectral detail. The value of fundamental frequency (F0) for singing has also been highlighted: RMSSinger employs diffusion‑based pitch modeling to enhance naturalness~\cite{he2023rmssinger}, and SiFiSinger~\cite{cui2024sifisinger} adds a source module that produces F0‑controlled excitation signals for finer pitch control. Moreover, by consistency models~\cite{song2023consistency}, researchers also propose high-efficiency and high-quality SVS solutions~\cite{ye2023comospeech,lu2024comosvc}.

\paragraph{Controllable Synthesis} Beyond high‑fidelity audio, contemporary SVS systems are expected to afford fine‑grained control over timbre, singing techniques, and singing style, trying to achieve a comprehensive improvement in quality and controllability. At the timbre level, MuSE‑SVS~\cite{kim2023muse} offers a multi‑singer, emotion‑aware synthesizer that enables timbre selection. To manipulate expressive prosody,~\cite{song2022singing} introduces a DL-based SVS model that controls multiple aspects of vibrato. For technique controllability, SinTechSVS~\cite{zhao2024sintechsvs} integrates an attention‑based local‑score module to model singing techniques. The emergence of LLM opens a new frontier in prompt‑based conditioning. PromptSinger~\cite{wang2024prompt} uses natural‑language prompts to steer timbre, emotion, and loudness, while TechSinger~\cite{guo2025techsinger} extends this paradigm to precisely control techniques with classifier free guidance ~\cite{ho2022classifier}. Notably, the expressive diversity of singing has motivated style‑specific generation models. Examples include FreeStyler for rap generation~\cite{ning2025drop} and SongSong for classical art‑song synthesis~\cite{hu2025songsong}. Future work is poised to explore multi‑modal prompt‑driven, multi‑dimensional control for richer user interaction.

\paragraph{Singing Style Transfer} Audio prompt conveys richer acoustic detail and more distinctive stylistic cues than textual information. Consequently, singing‑style transfer and voice‑conversion techniques have become pivotal research directions. The pioneering work in this field~\cite{shen2023naturalspeech,zhang2024stylesinger} demonstrates that residual vector quantization(RVQ) effectively captures diverse stylistic attributes. While TCSinger replaces RVQ with clustering vector quantization(CVQ) for more stable style compression and augments the system with an LM to jointly model prosody and style~\cite{zhang2024tcsinger}. ExpressiveSinger~\cite{dai2024expressivesinger} further enhances expressiveness by cascading diffusion‑based control modules.
Moreover, due to the scarcity of singing data and the high requirements for singers, speech‑to‑singing(STS)~\cite{li2023alignsts} paradigm has emerged, aiming to improve both musicality and style fidelity~\cite{dai2025everyone}. The request for broader personalization also spurs unified frameworks: TCSinger2 unifies controllable synthesis and style transfer via contrastive learning and a mixture of experts (MOE) architecture~\cite{zhang2025tcsinger}. Developing more customized and adaptable generation schemes remains a fruitful avenue for future research.

\paragraph{Text-to-Song Generation}
Song generation has also remained a sustained focus for audio researchers. One of the earliest works is JukeBox~\cite{dhariwal2020jukebox}. With the advance of SVS and music generation~\cite{donahue2023singsong}, a natural idea is to design cascaded frameworks. Prior works~\cite{hong2024text,li2024accompanied} decomposes song synthesis into two steps: (i) SVS and (ii) vocal-to-accompaniment generation. Versband~\cite{zhang2025versatile}further enhances this pipeline by MOE to improve controllability and vocal–accompaniment alignment.
Moreover, several studies have explored end-to-end generation. SongGen~\cite{liu2025songgen} employs a single-stage auto-regressive Transformer for text-to-song generation. Meanwhile, built upon the LLaMA2~\cite{touvron2023llama}, YuE~\cite{yuan2025yue} introduces structural progressive conditioning, addressing the challenge of long-form music generation. 
In addition to the auto-regressive paradigm, DiffRhythm~\cite{ning2025diffrhythm} and MuDiT~\cite{wang2024mudit} explore DiT~\cite{peebles2023scalable} backbones for non-autoregressive song generation, while Levo~\cite{lei2025levo} further introduces a mixed‑token strategy together with direct preference optimization~\cite{wallace2024diffusion}, achieving multi‑preference alignment and improving musicality.
These developments underscore the growing potential of end-to-end models in song generation.

\section{Architecture}\label{sec:arch}

\begin{figure*}[ht]
\centering
\includegraphics[width=0.96\linewidth]{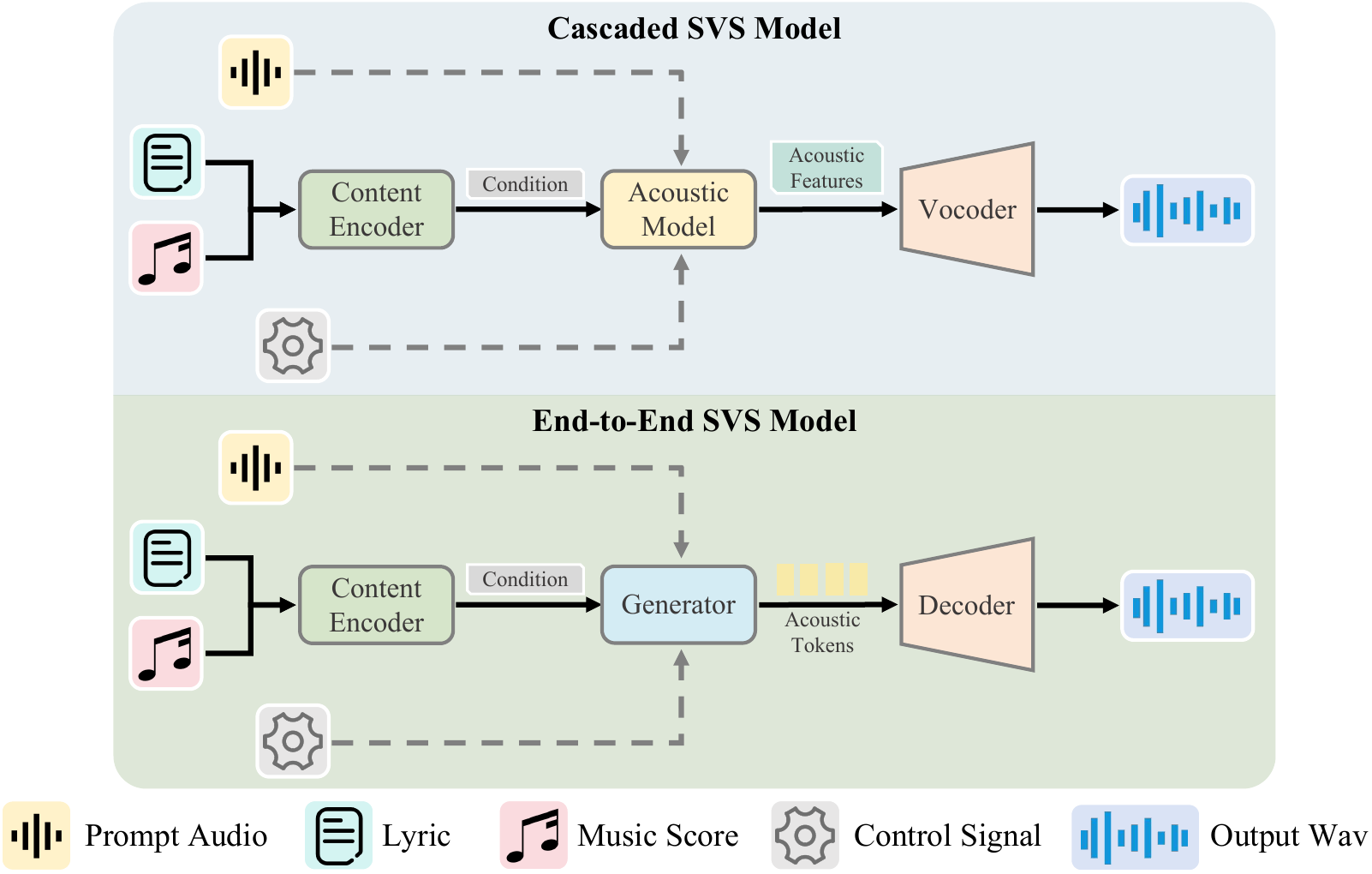}
\caption{We categorize SVS models into two paradigms, cascaded and end-to-end approaches. A system is end-to-end if it outputs waveforms directly, without intermediate modules or hand-crafted interfaces. Thus \textbf{requiring a vocoder implies a cascaded design}. Dashed lines indicate optional process.}
\label{fig:arch}
\end{figure*}
\vspace{-7pt}

In singing synthesis, the fundamental challenge is delivering stable, high‑quality outputs despite the one‑to‑many mapping from score to waveform. As shown in Figure~\ref{fig:arch}, we therefore classify SVS models by whether they employ a vocoder in waveform generation: cascaded and end-to-end frameworks. 

\subsection{Cascaded SVS Systems}

\paragraph{Acoustic Model} In cascaded models, the acoustic model manages the "music score $\to$ spectral feature" mapping. Great progress has been made in the network architecture for SVS. Inspired by Transformer-based TTS models, early DL-based SVS frameworks such as HiFiSinger~\cite{chen2020hifisinger} adopts non-autoregressive acoustic models, while ByteSing~\cite{gu2021bytesing} is designed based on an autoregressive Tacotron-like~\cite{wang2017tacotron} architecture. DeepSinger~\cite{ren2020deepsinger} propose a feed-forward Transformer-based singing model while using a large amount of internet singing data for training. WeSinger~\cite{zhang2022wesinger} introduces data augmentation methods and replaces the commonly used mel-spectrogram features with LPC features to improve accuracy. An adversarial multi-task learning framework~\cite{kim2022adversarial} is also novelly introduced to disentangle timbre and pitch features in singing. Furthermore, diffusion models~\cite{ho2020denoising}, as a flexible and interpretable generative framework, are also emerging in the audio modality. DiffSinger~\cite{liu2022diffsinger} employs a conditional diffusion process that, guided by score information, reconstructs target Mel‑spectrograms from Gaussian noise with high fidelity. RMSSinger~\cite{he2023rmssinger} further introduces a diffusion‑based pitch predictor to yield more accurate and controllable F0 trajectories. These approaches provide new avenues for improving the naturalness and expressiveness of synthesized singing. More recently, flow matching has emerged as a prominent generative paradigm; flow‑matching acoustic models like TechSinger~\cite{guo2025techsinger} offer faster and more stable generation while maintaining quality.

\paragraph{Vocoder} Before the widespread adoption of deep learning, vocoders were typically of two types: (i) algorithmic reconstruction, such as the Griffin–Lim algorithm; and (ii) parametric reconstruction, exemplified by WORLD~\cite{morise2016world}. Neural vocoders have since become prevalent due to their superior reconstruction quality and strong generalization. WaveRNN~\cite{kalchbrenner2018efficient} is a commonly used autoregressive vocoder in SVS systems. Non‑autoregressive alternatives include teacher–student frameworks~\cite{oord2018parallel} and diffusion‑based methods~\cite{chen2020wavegrad}. At present, the most widely adopted neural vocoders are GAN‑based, with HiFi‑GAN~\cite{kong2020hifi} and BigVGAN~\cite{lee2022bigvgan} demonstrating strong performance in audio reconstruction. Leveraging the unique characteristics of singing, researchers have also developed task‑specific neural vocoders tailored for SVS, enabling expressive reconstruction~\cite{huang2021multi}.

\subsection{End-to-End SVS Systems}
To mitigate error accumulation and train–test domain mismatch inherent to cascaded pipelines, researchers have advanced end‑to‑end SVS models. VISinger~\cite{zhang2022visinger} first brought the VITS architecture~\cite{kim2021vits} to SVS, delivering a fully end‑to‑end system. VISinger2~\cite{zhang2022visinger2} further integrates DSP techniques and raises the sampling rate to substantially improve acoustic fidelity. SiFiSinger~\cite{cui2024sifisinger}, an extension of VISinger, introduces a source module that generates F0‑controlled excitation signals for enhanced pitch control. UniSyn~\cite{lei2023unisyn} adopts a multi‑conditioned VAE~\cite{kingma2013auto} to provide flexible control over audio attributes and singing style. CSSinger~\cite{cui2025cssinger} designs a semi‑streaming framework that leverages latent representations within a VAE to achieve fully end‑to‑end streaming synthesis, improving quality while reducing latency.
Besides, inspired by LLMs, researchers have also explored neural‑codec–based high-fidelity SVS schemes~\cite{hwang2025hiddensinger,wu2024toksing}.

\section{SVS Representations and Control}\label{sec:modeling}

\subsection{Representations and Modeling of SVS}

Representations are central to SVS, determining how inputs are interpreted and how vocals are ultimately rendered. Required dimensions span content (lyrics, score), acoustic information and semantic embeddings. In this chapter, we survey commonly used representation and modeling choices for both content and audio signals in SVS systems.

\paragraph{Singing's Content Representation} Content representations focus on the question of \textit{"what to sing"} and \textit{"when to sing"}, mapping lyrics and score onto time, and providing the backbone for SVS stability. A common pipeline uses G2P-derived phoneme sequences\footnote{\url{https://github.com/Kyubyong/g2p}} fused with score cues (MIDI pitch, note duration/boundaries)~\cite{lu2020xiaoicesing, liu2022diffsinger}. To address the one‑to‑many rhythm mapping and melisma, systems rely on: (i) external forced alignment for phoneme/syllable durations~\cite{he2022pama}; (ii) learnable monotonic alignment with stochastic duration prediction to model rhythmic uncertainty~\cite{zhang2022visinger}; and (iii) learnable up‑sampling or a length regulator to expand token‑level states to frames~\cite{he2023rmssinger,zhang2024stylesinger}. Despite wide adoption, studies have highlighted limitations in robustness and naturalness~\cite{jiang2025megatts}. Consequently, recent systems refine alignment modeling, for example by introducing RL–based optimization to improve perceptual objectives~\cite{li2025dmospeech}, and masked‑token representations to stabilize monotonic alignments~\cite{zhang2025tcsinger}.

\paragraph{Singing's Acoustic Representation}  
Acoustic representations determine “who sings” and largely drive perceived quality and naturalness. Cascaded SVS predicts hand-crafted targets like Mel, F0, and UV, because they are stable and easy to optimize~\cite{liu2022diffsinger,zhang2022wesinger,he2023rmssinger}.
Recent work adopts learned latents to capture long‑range structure and micro‑textures. Two main lines are: (i) neural‑codec tokens with RVQ, providing compact, robust discrete acoustic units~\cite{zeghidour2021soundstream}; and (ii) continuous latents from VAE/RQ‑VAE encoders, typically arranged in multi‑band or multi-scale hierarchies to balance timbral detail and global coherence \cite{lee2022autoregressive}.
Among discrete methods, HiddenSinger discretizes acoustics with stacked RVQ blocks \cite{hwang2025hiddensinger}, while TokSing fuses VQ‑VAE, RVQ, and clustered SSL embeddings to construct tokens \cite{wu2024toksing,van2017neural}.
Meanwhile, continuous latents are usually adopted to couple high‑level conditioning with waveform rendering.  VITS-like~\cite{kim2021vits} SVS systems use VAE latents with a learned prior to model frame‑level variability while keeping alignment monotonic~\cite{zhang2022visinger2}. UniSyn~\cite{lei2023unisyn} further introduces a multi‑conditional VAE that factorizes speaker and style subspaces. More recently, contrastive learning have been coupled with VAEs to learn style‑aware acoustic tokens~\cite{zhang2025tcsinger}.

\paragraph{Singing's Semantic Representation} Semantic representations target content and expressivity, shaping correctness and control, and primarily focus on the question of “how to sing.”
Early work on semantic representations used classical ML (SVMs, HMMs) with hand‑crafted spectral and statistical features. Deep learning then popularized CNN/RNN extractors for speech and singing semantics \cite{shirian2021compact}.
Pretrained speech encoders have since delivered further breakthroughs. For emotion recognition, wav2vec 2.0~\cite{baevski2020wav2vec} employs masked latent modeling with quantized contrastive learning, setting strong baselines for speaker and emotion recognition \cite{huang2022generspeech}. HuBERT~\cite{hsu2021hubert} couples offline clustering targets with masked‑region prediction to learn joint acoustic–linguistic representations; its hierarchical attention captures prosodic cues linked to emotion. WavLM~\cite{chen2022wavlm} jointly optimizes masked‑speech prediction and denoising, further improving discriminability. And a systematic evaluation~\cite{atmaja2022evaluating} reports strong performance from both HuBERT and WavLM.
Moreover, given the expressive power of language, LLM‑based interfaces are emerging for complex singing semantics. For instance, SeCap~\cite{xu2024secap} uses a Q‑Former bridge to convert speech into style‑aware tokens for LLaMA \cite{touvron2023llama}, and this semantic modeling solution may also be a beneficial exploration direction.

\subsection{Control Techniques for Generation}

Style control in audio generation mainly follows two routes: audio‑based transfer and text‑based control. Audio‑driven transfer is especially popular because it avoids extra annotations, and recently style control via multi-modal prompting has gradually become a research hotspot. These approaches have matured into several representative pathways, delivering notable advances in both TTS and SVS~\cite{jiang2023mega,zhao2024sintechsvs}.

\paragraph{Audio-based Style Transfer}
Within classical autoregressive TTS, Wang et al.~\cite{wang2018style} introduces Global Style Tokens (GST), establishing the paradigm for disentangled, controllable prosody. Attentron~\cite{choi2020attentron} further leverages an attention mechanism to extract prosodic cues, enabling fine‑grained transfer. Building on this line, Li et al.~\cite{li2021towards} designs a multi‑scale reference encoder that captures phoneme-level style, achieving precise control.
Spurred by advances in non‑autoregressive audio generation~\cite{lam2021bilateral}, research has explored new style‑transfer paradigms. Styler~\cite{lee2021styler} factorizes style into disentangled components and Mega‑TTS~\cite{jiang2024mega} models prosody with a latent LM, learning distributions over rhythmic and stylistic patterns to support style transfer. HierSpeech line~\cite{lee2022hierspeech,lee2023hierspeech++} conditions jointly on text and audio prompts to generate pitch, providing control over prosodic style. In SVS, researchers advance beyond GST by introducing style adaptors and adaptive decoders, enabling more versatile transfer~\cite{zhang2024stylesinger}. Moreover, in singing voice conversion, a well-established paradigm disentangles the reference audio into multiple factors (content, rhythm, timbre, etc.) and then fuses them to achieve zero-shot transfer~\cite{li2023alignsts,dai2025everyone}.

\paragraph{Text-based Style Control}
A natural route to precise style control is to inject control signals by concatenation or addition. PromptSinger~\cite{wang2024prompt} concatenates style features within a multi‑scale Transformer to steer timbre, emotion, and loudness. SinTechSVS~\cite{zhao2024sintechsvs} follows a similar idea, introducing an attention‑based local‑score module to enhance controllability of specific singing techniques. A direct extension is cross‑attention, and Lyth et al.~\cite{lyth2024natural} model speaker style and emotion in this way, removing dependence on reference audio. In parallel, many SVS systems adopt adaptive normalization~\cite{huang2017arbitrary,peebles2023scalable} to modulate intermediate activations with style codes, providing lightweight, continuous control that composes well with attention‑based conditioning~\cite{zhang2025versatile}. Moreover, the widespread use of conditional generators in SVS naturally enables CFG–based control~\cite{ho2022classifier}, allowing users to adjust the strength of style conditioning at inference~\cite{guo2025techsinger}. Notably, recent work also introduces MoE controllers: by learning routing policies that select specialized generation experts, these systems achieve high‑quality style control while preserving robustness and efficiency~\cite{zhang2025tcsinger}.

\section{Resources in SVS Models}\label{sec:resources}

\begin{table*}[ht]
    \centering
    \small
    \setlength{\tabcolsep}{10.5pt}
    \caption{Overview of publicly available singing voice datasets. \textbf{Lang} denotes the number of supported languages. \textbf{Dur. (hour)} refers to the total annotated duration. \textbf{Score} indicates whether musical score or note-level information is provided. \textbf{Align} shows the availability of precise text-audio alignment. \textbf{Style} indicates whether vocal style annotations are included.}
    \scalebox{0.8}{
    \begin{tabular}{c|c|c|c|c|c|c|c}
    \toprule
    \textbf{Corpus} & \textbf{Lang} & \textbf{Song} & \textbf{Singer} & \textbf{Dur. (hour)} & \textbf{Score} & \textbf{Align} & \textbf{Style} \\
    \midrule
    VocalSet~\cite{wilkins2018vocalset} & 1 & 3(mainly vocalise) & 20 & 10.1 & \XSolidBrush & \XSolidBrush & \XSolidBrush \\
    PJS~\cite{koguchi2020pjs} & 1 & 100(sentences) & 1 & 0.5 & Both & \XSolidBrush & \Checkmark \\
    MIR-1K~\cite{hsu2009improvement} & 1 & 110 & 19 & 2.2 & \XSolidBrush & Voiced type & \XSolidBrush \\
    NUS-48E~\cite{duan2013nus} & 1 & 20 & 12 & 2.8 & \XSolidBrush & \Checkmark & \XSolidBrush \\
    KVT~\cite{kim2020semantic} & 1 & 466 & 114 & 18.9 & \XSolidBrush & \XSolidBrush & \Checkmark \\
    CSD~\cite{choi2020children} & 2 & 100 & 1 & 4.9 & MIDI & \Checkmark & \XSolidBrush \\
    NHSS~\cite{sharma2021nhss} & 1 & 20 & 10 & 4.8 & \XSolidBrush & word & \XSolidBrush \\
    OpenSinger~\cite{huang2021multi} & 1 & 1146& 66& 50 & \XSolidBrush & \XSolidBrush & \XSolidBrush \\
    Tohoku Kiritan~\cite{ogawa2021tohoku} & 1 & 50 & 1 & 3.5 & Score & \Checkmark & \XSolidBrush \\
    PopCS~\cite{liu2022diffsinger} & 1 & 117 & 1& 5.9 & \XSolidBrush & \XSolidBrush & \XSolidBrush \\
    M4Singer~\cite{zhang2022m4singer} & 1 & 700 & 20 & 29.8 & MIDI & \Checkmark & \XSolidBrush \\
    PopBuTFy~\cite{liu2022learning} & 2 & 542 & 34 & 50.8 & \XSolidBrush & \XSolidBrush & \Checkmark \\
    Opencpop~\cite{wang2022opencpop} & 1 & 100 & 1 & 5.3 & MIDI & \Checkmark & \XSolidBrush \\
    SingStyle111~\cite{dai2023singstyle111} & 3 & 111 & 8 & 12.8 & Both & \Checkmark & \Checkmark \\
    GTSinger~\cite{zhang2024gtsinger} & 9 & 1366 & 20 & 80.6 & Score & \Checkmark & \Checkmark \\
    ACE-Opencpop~\cite{shi2024singing} & 1& 100& 30& 4.3& MIDI& \Checkmark&\XSolidBrush\\
    ACE-KiSing~\cite{shi2024singing} & 2& 23& 34& 1.0& MIDI& \Checkmark&\XSolidBrush\\
    \bottomrule
    \end{tabular}}
    \label{tab:dataset}
\end{table*}
\vspace{-5pt}

\subsection{Open-Source Datasets}

High-quality datasets are the foundation of effective singing voice synthesis (SVS) systems. Compared with speech synthesis, SVS demands higher data quality and finer‑grained annotations to capture singing’s intrinsic complexity, making dataset collection substantially more challenging. Several singing corpora have been released to alleviate this data bottleneck, as summarized in Table~\ref{tab:dataset}.
VocalSet \cite{wilkins2018vocalset}, built for waveform‑concatenative synthesis, mainly records isolated phonemes and is thus unsuitable for natural SVS. For DL‑based SVS, MIR‑1K~\cite{hsu2009improvement}, PopBuTFy~\cite{liu2022learning} and NHSS ~\cite{sharma2021nhss} examine speech–singing relations, while NUS‑48E~\cite{duan2013nus} introduces phoneme‑level alignments, motivating later Mandarin corpora~\cite{huang2021multi,liu2022diffsinger}. Recent datasets add musical‑score annotations (pitch, note boundaries, duration, meter) to improve musicality in synthesis \cite{wang2022opencpop,choi2020children,koguchi2020pjs,ogawa2021tohoku}.
With rising demand for customized vocals, M4Singer \cite{zhang2022m4singer} and KVT \cite{kim2020semantic} introduce substantial diversity in timbre and style, respectively. The most customization‑oriented open corpora to date are \textbf{SingStyle111} \cite{dai2023singstyle111} and \textbf{GTSinger} \cite{zhang2024gtsinger}, which provide rich annotations alongside multilingual recordings. Additionally, ACE‑Opencpop and ACE‑KiSing \cite{shi2024singing} are synthesizer‑generated corpora, offering a complementary avenue for data augmentation.

\subsection{Annotation Tools for Singing Data}

\begin{figure}[ht]
\centering
\includegraphics[width=\linewidth]{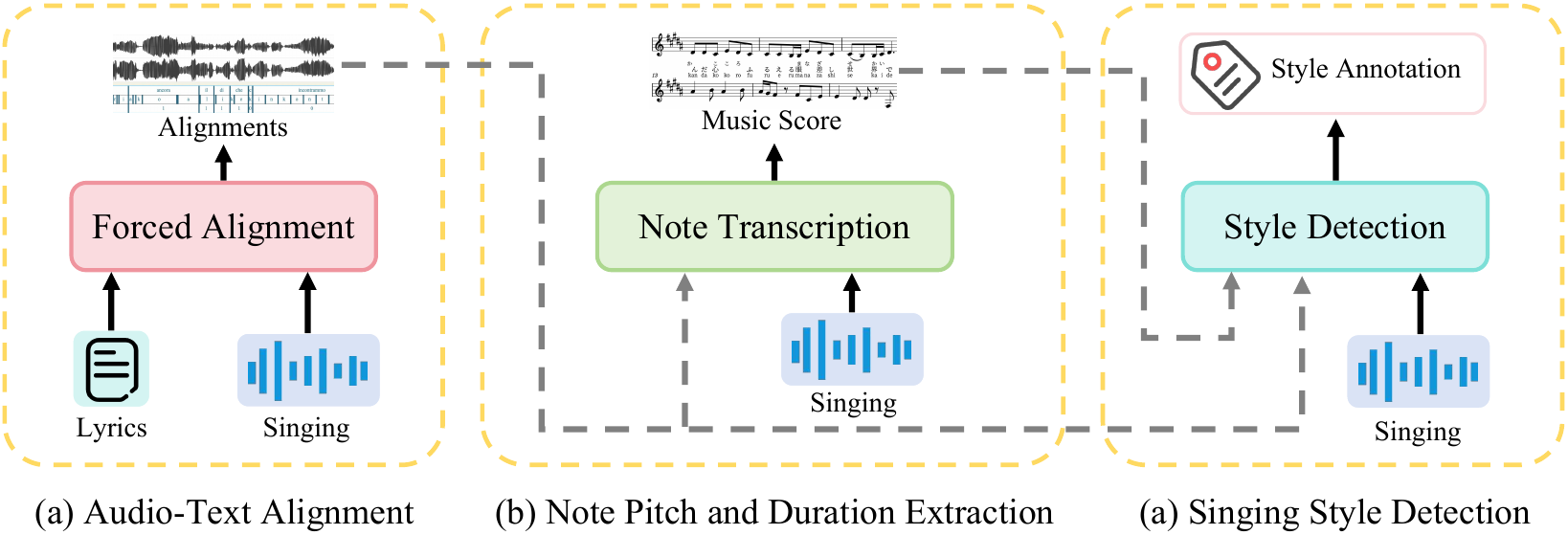}
\caption{
Three data annotation tasks required by the singing voice synthesis system.
}
\label{fig:annotation}
\end{figure}
\vspace{-5pt}

Training data for SVS should at minimum include lyric text, phoneme durations, and note information. The annotation tasks required by SVS is shown in Figure~\ref{fig:annotation}. For alignment, Praat\footnote{\url{https://www.fon.hum.uva.nl/praat/}} is the de‑facto tool for manual labeling; while MFA is an HMM-based automatic tool, which is widely and effectively used in clean vocals~\cite{mcauliffe2017montreal}; a CTC‑loss singing aligner, SOFA\footnote{\url{https://github.com/qiuqiao/SOFA}}, further builds on MFA. For vocal‑note transcription, Parselmouth\footnote{\url{https://github.com/YannickJadoul/Parselmouth}} provides convenient pitch extraction interfaces ~\cite{ren2020deepsinger}. More recent learning‑based systems achieve higher accuracy: ROSVOT leverages a Conformer~\cite{gulati2020conformer} backbone for robust pitch and duration estimation~\cite{li2024robust}, while MusicYOLO~\cite{wang2022musicyolo} adapts the YOLO detection framework to jointly detect note pitch and duration. As for style annotations, the principal methods are outlined in Section~\ref{sec:modeling}. Notably, recent work proposes end‑to‑end, multi‑task annotation frameworks~\cite{guo2025stars}, offering a more efficient route to automatic singing labeling.

\subsection{Evaluations for SVS}

Singing is intrinsically multifaceted, encompassing pitch accuracy, timbre similarity, naturalness, expressiveness, and technique, which makes evaluation both challenging and essential~\cite{cho2021survey}. This section details a range of evaluation metrics, re-categorized by the specific attribute they assess: accuracy, expressiveness \& naturalness, timbral quality , similarity, controllability.

\paragraph{Accuracy}
Accuracy metrics assess fidelity to score and lyrics—the basis of intelligible, musically correct SVS. Lyric accuracy, the most basic requirement, typically assessed using character error rate (CER), as in TTS~\cite{du2024cosyvoice,huang2023singing}. Pitch accuracy measures the conformity of the synthesized fundamental frequency (F0) contour to the target pitch derived from a musical score or a reference recording and it is a cornerstone metric in many SVS challenges (e.g., SVS-VC~\cite{huang2023singing}). Common criteria include F0 Frame Error (FFE)~\cite{zhang2024tcsinger} and Root Mean Square Error (RMSE)~\cite{zhang2022visinger2}, and some studies additionally provide the linear correlation between ground-truth and synthesized F0 contours~\cite{zhang2022wesinger}. What’s more, as a critical dimension for rhythmic integrity and intelligibility, duration and rhythm accuracy are commonly evaluated by Duration RMSE/MAE (the error between predicted and ground-truth phoneme durations) and Duration Prediction Accuracy.

\paragraph{Expressiveness and Naturalness}
Expressiveness metrics quantify artistic, emotional, and nuanced aspects beyond basic accuracy. Subjectively, Mean Opinion Score (MOS) remains the gold standard for perceptual quality~\cite{zhang2022m4singer}. Objectively, early work often sought to model subjective judgments. For example, SingMOS~\cite{tang2024singmos,tang2025singmos} introduces a dedicated dataset for singing MOS prediction, and many studies trained neural predictors on this resource; the VoiceMOS challenges~\cite{cooper2023voicemos,huang2024voicemos} also include tracks specifically targeting the prediction of subjective ratings for SVS and SVC. An emerging direction is using MLLM~\cite{chen2024mllm} to evaluate expressiveness. For instance, an RL agent can be trained to optimize a singing model toward a perceptual objective, such as maximizing a predicted emotion score, and the resulting reward can serve as an objective proxy for that expressive dimension~\cite{lei2025levo,bai2025dragon}. Another practical strategy is to leverage Singing Voice Deepfake Detection to assess the authenticity of vocals~\cite{zhang2024svdd}.

\paragraph{Sound Quality}

Noise- and artifact-free singing is a fundamental requirement for SVS. Early audio quality evaluation practices in SVS were adopted from TTS, including the use of PESQ~\cite{rix2001perceptual} for objective perceptual quality estimation and SNR~\cite{tandra2008snr} for quantifying the noise proportion in synthesized singing. Other studies quantify timbral differences between synthesized and reference audio by comparing spectral representations, employing metrics such as Mel-Cepstral Distortion (MCD)~\cite{zhang2024tcsinger} and alternatives based on Bark-frequency cepstral coefficients (BFCC)~\cite{zhang2022wesinger}.

\paragraph{Similarity}

Similarity metrics are crucial for tasks such as singing voice conversion (SVC) and voice cloning, assessing how well the synthesized voice matches a target singer’s identity or style. The most reliable approach remains subjective evaluation, exemplified by the Similarity Mean Opinion Score (MOS-S)~\cite{zhang2024stylesinger}, which uses a 5-point scale to rate the similarity between reference and synthesized audio. Preference-based protocols such as the AXY Test~\cite{skerry2018towards} ask listeners to choose, from two synthesized samples, the one closer to the reference, enabling robust cross-system comparison; this protocol is widely used in evaluations like the Singing Voice Conversion Challenge. On the objective side, Speaker Embedding Cosine Similarity (COS) is commonly employed, where SSL encoders~\cite{chen2022wavlm} extract speaker embeddings (e.g., x-vectors~\cite{snyder2018x}, d-vectors) and cosine similarity between reference and generated embeddings serves as a proxy for the model’s transfer capability.

\paragraph{Controllability}

Controllability metrics evaluate a model’s ability to precisely manipulate specified singing-voice attributes in response to user inputs. Control MOS (MOS-C)~\cite{zhang2024gtsinger} is a subjective test in which listeners judge how well the model follows a given instruction, such as altering style, emotion, or a singing technique. On the objective side, evaluation typically relies on attribute prediction models; for example, an emotion recognition model~\cite{ma2024emotion2vec} can infer the emotion of the synthesized audio, which is then compared with the input control signal. Common summary metrics include Accuracy and F1 score.

\section{Conclusion}

In this survey, we systematically review the recent research on singing voice synthesis based on deep learning from tasks and architectures to core technologies in SVS including singing modeling, control techniques and training strategies. We also collect and demonstrate datasets, annotation tools and evaluation benchmarks for SVS systems. 
Through reviewing recent progress, we hope to drive SVS development by clearer insights, gap identification, and ideas for more expressive synthesis.

\clearpage

\section*{Limitations}
Recent advances in generative models (e.g. GANs~\cite{goodfellow2014generative}, Transformers~\cite{vaswani2017attention}, diffusion models~\cite{ho2020denoising,rombach2022high,peebles2023scalable}, and flow‑matching frameworks~\cite{liu2022flow,lipman2022flow} have accelerated progress in singing voice synthesis. \textbf{Consequently, this survey mainly focuses on deep‑learning‑based SVS researches}. Relative to traditional methods such as waveform concatenation~\cite{kenmochi2007vocaloid} and statistical parametric synthesis~\cite{saino2006hmm}, modern deep models yield markedly better voice quality and naturalness, while affording superior controllability and generalization. For the above considerations, we don't introduce the paradigms of traditional singing voice synthesis at length.

\section*{Ethical Considerations}

Although this survey itself raises no immediate ethical concerns, two potential risks must be addressed when applying the reviewed methods. (1) \textbf{Data licensing.} Users must respect the licenses of public corpora and obtain explicit permission before crawling or repurposing web‑hosted singing recordings. (2) \textbf{Misuse of generative models.} Modern SVS systems can convincingly imitate a singer’s timbre and style; without safeguards, they may be used for unauthorized voice‑over, infringing intellectual‑property and personality rights. Practitioners should comply with model developers’ usage policies and local regulations, and future research should explore mitigation measures like voice‑print watermarking to protect singers’ privacy and provenance.

\section*{Acknowledgements}

This work was supported by National Natural Science Foundation of China under Grant No.U24A20326.

\bibliography{custom}

\clearpage
\appendix

\section{Training and Inference Strategies}
\label{sec:train}

\subsection{Training Data Augmentation}

Public singing datasets often suffer from limited scale, uneven recording quality, and inconsistent annotations, creating a data bottleneck that poses substantial challenges to training stable and robust SVS systems~\cite{zhang2024gtsinger}. Data augmentation seeks to expand acoustic‑condition diversity while preserving score consistency. Common strategies include: (a) \textbf{Score‑aware pitch/tempo transforms}. Specifically, do semitone transposition and BPM‑proportional time‑stretching, with synchronized updates to MIDI, F0, and durations, which enlarges timbre/prosody coverage while preserving musical structure~\cite{guo2022singaug,zhang2021pdaugment}. (b) \textbf{Spectral perturbations}. The application of spectrogram perturbations, such as frequency/time masking and spectrogram-domain mixup, functions as an effective regularizer for the acoustic model and effectively improves robustness~\cite{park2019specaugment,kim2021specmix}. (c) \textbf{F0‑centric perturbations}. To add small vibratos and slightly shift UV boundary for training data could effectively improve cross‑domain generalization while keeping musicality~\cite{bai2024spa}. (d) \textbf{Speech infusion}. Incorporating speech data expands style and prosody coverage by exploiting shared human‑voice attributes, benefiting the controllability of SVS~\cite{wang2024prompt}. Together, these techniques alleviate data scarcity, improve robustness, and yield more controllable SVS models.

\subsection{Training Strategies}
Training strategies in SVS aim to balance quality, controllability, data efficiency, and inference cost. We summarize two complementary axes. 
\paragraph{Pre‑train \& Fine‑tune} Large‑scale self supervised encoders~\cite{baevski2020wav2vec,hsu2021hubert} and singing‑specific pre‑training~\cite{li2024self} provide robust features and transfer well to low‑resource singers and languages. The above training methods are generally based on the following steps. Firstly, freeze or partially fine‑tune the SSL backbone while training SVS heads; then, add PEFT for efficient style adaptation; and finally use domain adapters to bridge speech and singing. Such strategies are also applicable to latent diffusion backbones~\cite{liu2023audioldm}.
\paragraph{Multi-Stage Training} For cascaded systems, a typical schedule trains the acoustic model and duration/F0 (note) predictors first, followed by the vocoder; a subsequent joint fine‑tuning stage aligns the conditional distributions and mitigates error accumulation~\cite{chen2020hifisinger,liu2022diffsinger}.
For consistency‑model frameworks, lightweight high‑throughput synthesis is commonly achieved via teacher–student distillation, compressing sampling steps while preserving fidelity; this distillation is performed after training a high‑quality teacher (often a diffusion model) and serves as its downstream refinement~\cite{ye2023comospeech}.

\subsection{Inference Acceleration Method}

In addition to the consistency model, rapid progress in deep generative modeling~\cite{ho2020denoising,liu2022flow,he2025diffsr,huang2025psdiffusion} enable faster inference for diffusion-based and flow-matching SVS systems. For diffusion models, fast ODE/SDE solvers—notably DDIM~\cite{song2020denoising} and DPM‑Solver~\cite{lu2022dpm}—recast the reverse process as an ODE and use higher‑order, adaptive updates to achieve high fidelity with few steps. For flow matching, which learns an explicit velocity field from data to noise, acceleration is largely intrinsic: integrators provide short trajectories with minimal overhead~\cite{guo2025techsinger}. Beyond step‑size reduction, methods that smooth the vector field like Rectified Flow~\cite{liu2022flow} and OT‑Flow~\cite{onken2021ot} regularize the transport dynamics to yield straighter, more stable paths, thereby further cutting the number of required steps. 

\section{Discussions about Singing Voice Synthesis}

\subsection{Novel Singing Tasks}

As discussed in Section~\ref{sec:task}, recent SVS systems already achieve strong performance on core accuracy dimensions—e.g., lyric and pitch correctness~\cite{liu2022diffsinger,zhang2022visinger}. Nevertheless, a substantial gap remains between current outputs and human expectations~\cite{zhang2025tcsinger}. This gap stems in part from the community’s ambitions moving beyond merely “singing the notes correctly” toward richer objectives that encompass expressivity, style, and realism. Building on this, future SVS research can advance along two axes: (1) customization and (2) high expressivity. For \textbf{customization-oriented singing}, instruction following should be prioritized once basic audio quality is ensured. A practical route is to strengthen both generative capacity and cross-modal alignment—for example, by using MLLMs~\cite{comanici2025gemini} to encode heterogeneous conditioning into a unified representation, or by leveraging contrastive learning~\cite{elizalde2023clap} to build a task-specific, multimodal singing representation. Moreover, reinforcement learning has demonstrated improvements in instruction following across domains~\cite{liu2024deepseek,bai2025dragon,du2025cosyvoice}. For \textbf{high expressivity singing}, beyond fine-grained control over dimensions such as emotion and style~\cite{dai2024expressivesinger}, and beyond joint vocal–accompaniment generation~\cite{ning2025diffrhythm,liu2025songgen}, an equally promising direction is \textbf{immersive (spatial) singing synthesis}. While spatial audio generation has seen breakthroughs in speech and sound effects~\cite{liang2025binauralflow,sun2024both,zhang2025isdrama}, it remains underexplored for singing. Given its growing deployment in in-car and headphone scenarios~\cite{pan2025multimodal,zhu2025asaudio} and the recent release of recorded spatial-singing datasets~\cite{guo2025mrsaudio}, we anticipate rapid progress in immersive singing generation.

\subsection{Architectures of SVS}

The architectural discussion for SVS systems can be framed along two perspectives: (i) cascaded vs. end-to-end designs, and (ii) autoregressive(AR) vs. non-autoregressive(NAR) paradigms.

\paragraph{Cascaded vs. End-to-End} Evidently, end-to-end (single-stage) generation is a more promising direction, benefiting from direct modeling. In contrast to TTS, where end to end designs are widely adopted~\cite{zhang2025minimax,du2025cosyvoice}, the SVS community still features many strong cascaded systems~\cite{zhang2024tcsinger}. A likely reason is that cascaded pipelines rely on hand-crafted targets such as Mel spectrograms to regularize the acoustic stage, which adds auxiliary supervision and reduces the compression rate, thereby securing a higher performance floor, especially in low resource settings~\cite{wang2022opencpop}. By comparison, end to end systems offer a higher ceiling but face challenges, including greater data requirements, training instability, and heightened sensitivity to alignment and temporal modeling~\cite{peebles2023scalable}. Recent work often mitigates these issues by introducing neural codecs or continuous latents as intermediate representations to ease waveform prediction~\cite{wu2024toksing,hwang2025hiddensinger}. Looking ahead, end to end SVS can be advanced by reducing data and compute requirements and by incorporating reinforcement learning and contrastive learning to improve perceptual quality. These directions merit deeper investigation.

\paragraph{AR vs. NAR} Inspired by FastSpeech~\cite{ren2020fastspeech}, NAR SVS gained broad adoption in academia and industry due to its inference speed. However, the rise of long form generation introduces unavoidable compute overheads for non autoregressive models, and the progress of MLLMs together with the inherently sequential nature of singing has renewed interest in autoregressive modeling~\cite{liu2025songgen}. Autoregressive generation, in turn, can suffer from training instability and longer inference time~\cite{wang2017tacotron}. Hybrid approaches that combine autoregressive and non autoregressive modeling, such as DITAR~\cite{jia2025ditar}, as well as streaming inference~\cite{cui2025cssinger}, may provide practical avenues for SVS.

\subsection{Representation of SVS}

\paragraph{Continuous Representation vs. Discrete Representation}
There is no consensus on whether singing voice generation should adopt continuous or discrete representations. Discrete representations~\cite{zeghidour2021soundstream} align naturally with autoregressive and LLM next-token objectives, enabling long-context caching, streaming inference, instruction following, and precise control~\cite{hwang2025hiddensinger}; however, quantization may smooth out fine vocal details. Continuous representations preserve fine-grained acoustic textures and phase structure, yet they often rely on powerful decoders and are more sensitive to alignment~\cite{lei2023unisyn,kim2021vits}. A dual-track approach that combines discrete and continuous representations may offer the best balance among efficiency, controllability, and high fidelity for SVS.

\paragraph{Trade-off between Quality and Controllability} Experiences show that an overemphasis on control can degrade audio fidelity~\cite{sadat2024eliminating}. Moreover, achieving sufficient disentanglement of content, acoustic, and semantic factors, together with effective cross-lingual and cross-style transfer, remains challenging~\cite{jiang2025megatts}. Therefore, relying on a single encoder to extract control and transfer features is insufficient, and increasing guidance strength only at inference time does not yield genuine controllability~\cite{ho2022classifier}. To address these issues, future work can explore sparse architectures such as Mixture-of-Experts~\cite{zhou2022mixture} to enhance style control and transfer~\cite{zhang2025versatile}, and apply reinforcement learning to align model behavior with human preferences~\cite{wang2025notagen}.

\subsection{Data for SVS}

 A significant challenge for SVS is the scarcity and quality of training data~\cite{zhang2022m4singer}. In contrast to speech datasets, which exceed 100,000 hours, open-source singing data is extremely limited and plagued by long-tail distributions in style and language, as well as poor annotation quality (e.g., inaccurate lyric timestamps)~\cite{zhang2024gtsinger}. Consequently, for the SVS community, a more pragmatic strategy is to pursue large-scale, weakly-labeled data collection rather than investing heavily in small, perfectly annotated corpora~\cite{kong2020panns}. This points to two crucial long-term research directions: developing efficient data processing pipelines~\cite{he2024emilia} and designing effective self-supervised or weakly-supervised encoding schemes for singing~\cite{li2024self}.

\section{Potential Contribution of MLLMs to SVS Field}

Multimodal Large Language Models (MLLMs)~\cite{xu2025qwen3,he2025radarqa,wang2025gre} are increasingly shaping the landscape of Singing Voice Synthesis (SVS). To highlight their growing significance, we provide a systematic discussion of their contributions from six key perspectives. This section outlines how MLLMs are driving progress in SVS, offering both new methodological insights and practical advancements for the field.

\subsection{Data Captioning and Annotation}

One of the longstanding challenges in SVS is the scarcity of richly annotated singing datasets, especially those with semantic labels such as emotion, style, or techniques~\cite{guo2025stars,dai2023singstyle111}. MLLMs offer a powerful solution by enabling automated, high-level semantic annotation. For instance, Automatic Speech Recognition(ASR) systems like Whisper~\cite{bain2022whisperx} and FireRedASR~\cite{xu2025fireredasr} can first transcribe singing content, while advanced MLLMs (e.g., GPT-4o~\cite{achiam2023gpt}, Gemini 2.5 Pro~\cite{comanici2025gemini}) can then be prompted to analyze the transcribed lyrics, melody context, and acoustic features to generate descriptive captions~\cite{wang2025spark}. This capability significantly reduces manual annotation costs and enables the creation of semantically enriched datasets that support more expressive and controllable synthesis.

\subsection{Content Understanding and Generation}

Beyond annotation, MLLMs excel at deep semantic understanding and creative generation of musical content~\cite{lei2024songcreator,ding2024songcomposer}. They can assist in high-level music composition tasks such as lyric writing, melody suggestion, rhyme structuring, and even genre-aware song structuring~\cite{bai2024seed,yuan2025yue}. By conditioning on textual prompts, MLLMs can generate coherent and stylistically appropriate lyrics that align with intended emotional narratives. Moreover, when integrated with symbolic music models or score-based diffusion systems~\cite{wang2025notagen}, they enable end-to-end co-creation of lyrics and melodies, forming a crucial bridge between natural language semantics and musical structure.

\subsection{Expressiveness Prediction and Guidance}

A key frontier in SVS is achieving human-like expressiveness, with subtle variations in pitch, timing, dynamics, and timbre that convey emotion and artistry~\cite{zhang2024gtsinger,yu2024visinger2+}. MLLMs may contribute by modeling the pragmatic and contextual aspects of singing performance. Recent works such as Prompt-Singer~\cite{wang2024prompt} and TechSinger~\cite{guo2025techsinger} demonstrate how textual prompts describing singing styles or techniques can guide systems to produce more contextually appropriate and nuanced outputs. MLLMs serve as interpreters between human intentions and acoustic parameters, enabling intuitive control over expressiveness without requiring technical expertise in signal processing.

\subsection{Voice and Song Generation}

Thanks to their strong long-sequence modeling capabilities and emergent multimodal alignment, large models have become foundational in end-to-end singing voice and full-song generation. Systems like YuE~\cite{yuan2025yue}, Seed-Music~\cite{bai2024seed}, Suno~\footnote{\url{https://suno.com/}}, and Mureka~\footnote{\url{https://www.mureka.ai/}} leverage MLLM-like architectures to generate high-quality singing voices directly from text and melody inputs, often in a zero-shot or few-shot manner. These models capture complex dependencies across lyrics, rhythm, pitch, and prosody, producing musically coherent and emotionally engaging results. Furthermore, emerging "all-in-one" audio foundation models such as AudioGen-Omni~\cite{wang2025audiogen} and UniFlow-Audio~\cite{xu2025uniflow} push the boundary by supporting unified generation of speech, sound effects, and singing from diverse inputs, including text and video modalities, highlighting the potential of MLLMs as general-purpose audio creators.

\subsection{Cross-Modal Alignment}

MLLMs inherently promote better cross-modal alignment between text, music, and audio, ensuring semantic consistency across layers of expression~\cite{elizalde2023clap}. For example, if a lyric mentions “a storm is coming,” an MLLM-guided SVS system can adjust both the vocal intensity and background instrumentation to reflect tension or drama, thereby enhancing narrative coherence. This holistic integration of meaning across modalities is difficult to achieve with traditional pipeline systems but emerges naturally in MLLM-based frameworks due to their joint training on vast multimodal corpora.

\subsection{MLLMs as SVS Evaluation}

Beyond generation, MLLMs can serve as intelligent evaluators for singing voice synthesis. Traditional metrics and small-scale human tests often fail to capture semantic fidelity, emotional expression, or musical naturalness~\cite{anastassiou2024seed}. MLLMs, with their cross-modal understanding, can assess synthesized singing by answering targeted questions~\cite{zhou2025scientists}. Inspired by the “LLM-as-a-Judge” paradigm~\cite{chen2024mllm}, this approach has been already widely applied in TTS benchmarks~\cite{manku2025emergenttts,huang2025instructttseval} and shows strong potential for automating and scaling SVS evaluation with high correlation to human judgment. Moreover, MLLMs can generate interpretable feedback, offering actionable insights beyond scalar scores. This enables faster, more informative model iteration and paves the way toward standardized, explainable evaluation in future SVS research.

In summary, MLLMs are not merely auxiliary tools in SVS; they are becoming central enablers of semantic richness, expressivity, and accessibility in singing synthesis. Their integration marks a paradigm shift, from purely signal-driven systems to intention-aware, context-sensitive, and user-centered music generation platforms.

\end{document}